\documentclass[aps,prl,reprint,superscriptaddress]{revtex4-1}

\usepackage{amsmath}
\usepackage{amssymb}
\usepackage{graphicx}
\usepackage{graphbox}

\usepackage[unicode=true,
 bookmarks=true,bookmarksnumbered=false,bookmarksopen=false,
 breaklinks=false,pdfborder={0 0 1},backref=false,colorlinks=true]
 {hyperref}
\hypersetup{
 linkcolor=magenta, urlcolor=blue, citecolor=blue, pdfstartview={FitH}, unicode=true}

\usepackage{amsfonts}\usepackage{tabularx}\usepackage{dcolumn}\usepackage{bm}\usepackage{graphicx}\usepackage{epstopdf}\usepackage{xcolor}
\hypersetup{urlcolor=blue}
\usepackage{times}

\usepackage{float}
\makeatletter
\let\newfloat\newfloat@ltx
\makeatother

\usepackage{algorithm}
\usepackage{algorithmic}

\begin{document}

%\preprint{}

\title{Topological Quantum Compiling with Reinforcement Learning}


\author{Yuan-Hang Zhang}
%\thanks{}
\affiliation{Center for Quantum Information, IIIS, Tsinghua University, Beijing 100084, People’s Republic of China}
\affiliation{Department of Physics, University of California, San Diego, CA 92093, USA}

\author{Pei-Lin Zheng}
\affiliation{International Center for Quantum Materials, Peking University, Beijing 100871, China}
\affiliation{School of Physics, Peking University, Beijing 100871, China}

\author{Yi Zhang}
\email{frankzhangyi@gmail.com}
\affiliation{International Center for Quantum Materials, Peking University, Beijing 100871, China}
\affiliation{School of Physics, Peking University, Beijing 100871, China}

\author{Dong-Ling Deng}
\email{dldeng@tsinghua.edu.cn}
\affiliation{Center for Quantum Information, IIIS, Tsinghua University, Beijing 100084, People’s Republic of China}
\affiliation{Shanghai Qi Zhi Institute, 41th Floor, AI Tower, No. 701 Yunjin Road, Xuhui District, Shanghai 200232, China}

\date{\today}

\begin{abstract}
Quantum compiling, a process that decomposes the quantum algorithm into a series of hardware-compatible commands or elementary gates, is of fundamental importance for quantum computing.  We introduce an efficient algorithm based on deep reinforcement learning that compiles an arbitrary single-qubit gate into a sequence of elementary gates from a finite universal set. It generates near-optimal gate sequences with given accuracy and is generally applicable to various scenarios, independent of the hardware-feasible universal set and free from using ancillary qubits. For concreteness, we apply this algorithm to the case of topological compiling of Fibonacci anyons and obtain near-optimal braiding sequences for arbitrary single-qubit unitaries.  Our algorithm may carry over to other challenging quantum discrete problems, thus open up a new avenue for intriguing applications of deep learning in quantum physics.
\end{abstract}

%\keywords{}

\maketitle

\begin{figure}
\hspace*{-0.48\textwidth}
\includegraphics[width = 0.46\textwidth]{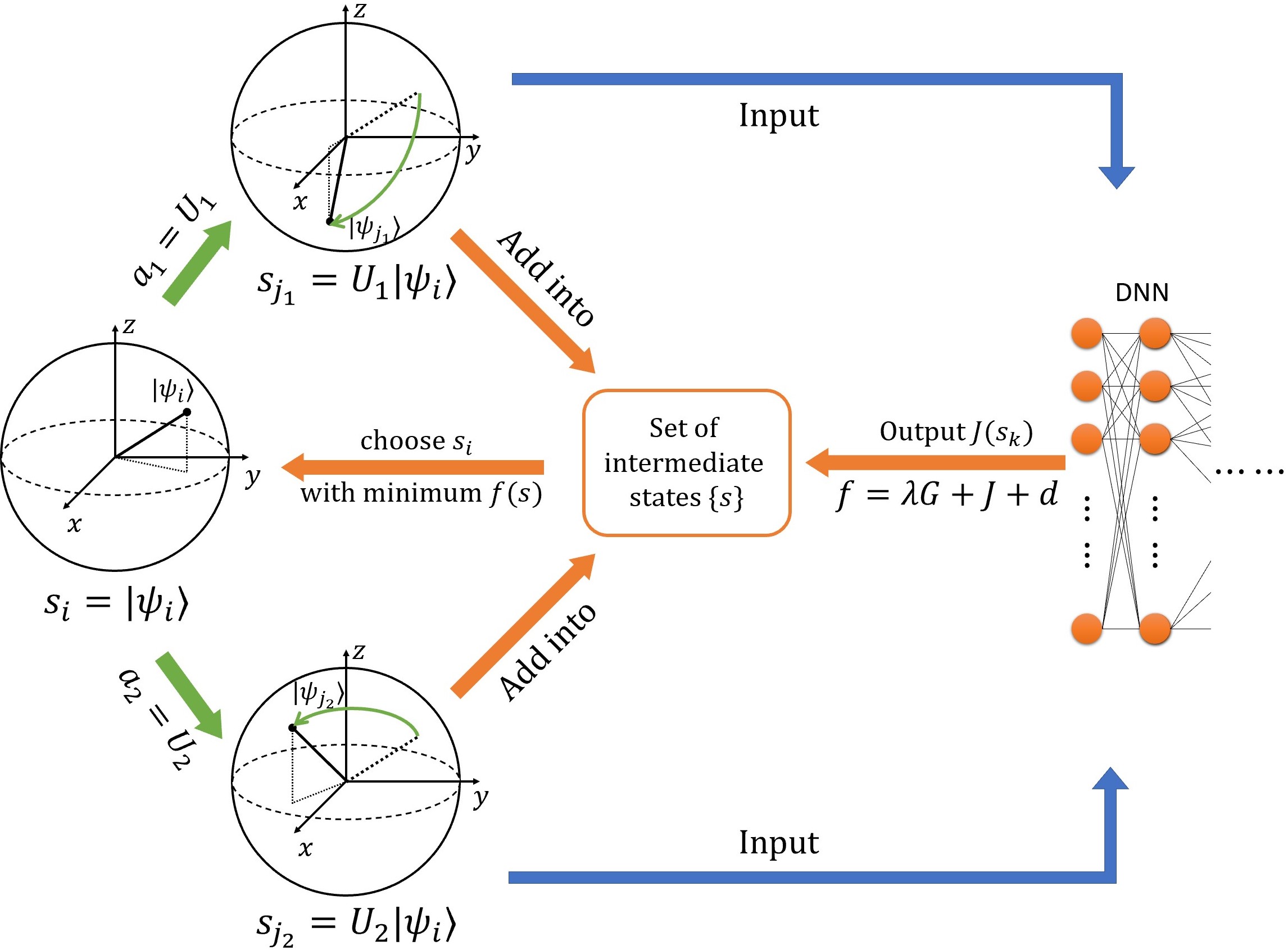}%
\caption{A schematic illustration of the algorithm: during the search, we start from the initial state $s_i=|\psi_i\rangle$, then iteratively pick the state with the minimum evaluation function $f(s)$ and evaluate all its successors with the DNN until we reach the target state (see \cite{SuppTQCviaRL}).\label{fig:algorithm schematic}}
\end{figure}

%\emph{Introduction.}\textemdash
To efficiently decompose unitaries into a sequence of elementary hardware-compatible quantum gates as short as possible is a crucial problem in a variety of quantum-information-processing tasks, such as quantum computing \cite{nielsen2002quantum} and quantum digital simulations \cite{Georgescu2014Quantum}. This problem becomes especially relevant for the noisy intermediate-scale quantum devices \cite{Preskill2018quantumcomputingin}, where the depth of the quantum circuits might be limited due to the inaccuracy of the possible elementary gates and quantum decoherence.  A number of notable algorithms have been proposed to compile single-qubit unitaries \cite{dawson2005solovay,Kitaev2002Classical,Jones2012Faster,Fowler2011Constructing,Bocharov2012Resource,Bocharov2013Efficient,Pham2013Optimization,Zhiyenbayev2018Quantum,kliuchnikov2013asymptotically,Selinger2013Quantum,Gosset2014An,Ross2016,heyfron2018efficient,Alam2019RL}. For instance, the Solovay-Kitaev algorithm runs in $O(\log ^{2.71}(1/\epsilon ))$ time and can output a sequence of $O(\log ^{3.97}(1/\epsilon )$ elementary gates that approximate the targeted unitary to precision  $\epsilon$ \cite{nielsen2002quantum, dawson2005solovay}. Other algorithms either exploit the specific structure of the Clifford$+T$ gate set \cite{kliuchnikov2013asymptotically,Selinger2013Quantum,Gosset2014An,Ross2016,heyfron2018efficient}, or utilize ancillary qubits \cite{Kitaev2002Classical,Jones2012Faster}, to further reduces the running time and length of the desired gate sequences. Each of these algorithms bears its pros and cons, and the choice depends on the specific problem. Here, inspired by the similarity between quantum compiling and solving Rubik's cube (see Table \ref{tab:comparison}), we introduce a novel algorithm based on deep reinforcement learning, which compiles single-qubit unitaries efficiently and is generally applicable to different scenarios (see Fig. \ref{fig:algorithm schematic} for an illustration).

\newcommand{\centered}[1]{\begin{tabular}{c} #1 \end{tabular}}

\begin{table}  %[H] add [H] placement to break table across pages
\begin{ruledtabular}
\begin{tabular}{ccc}
% note: the figures seem to misplace themselves without the hspace. I don't know why, but adding the hspace do solved the problem.
System &\hspace*{-0.16\textwidth}\includegraphics[align=c,width=0.16\textwidth]{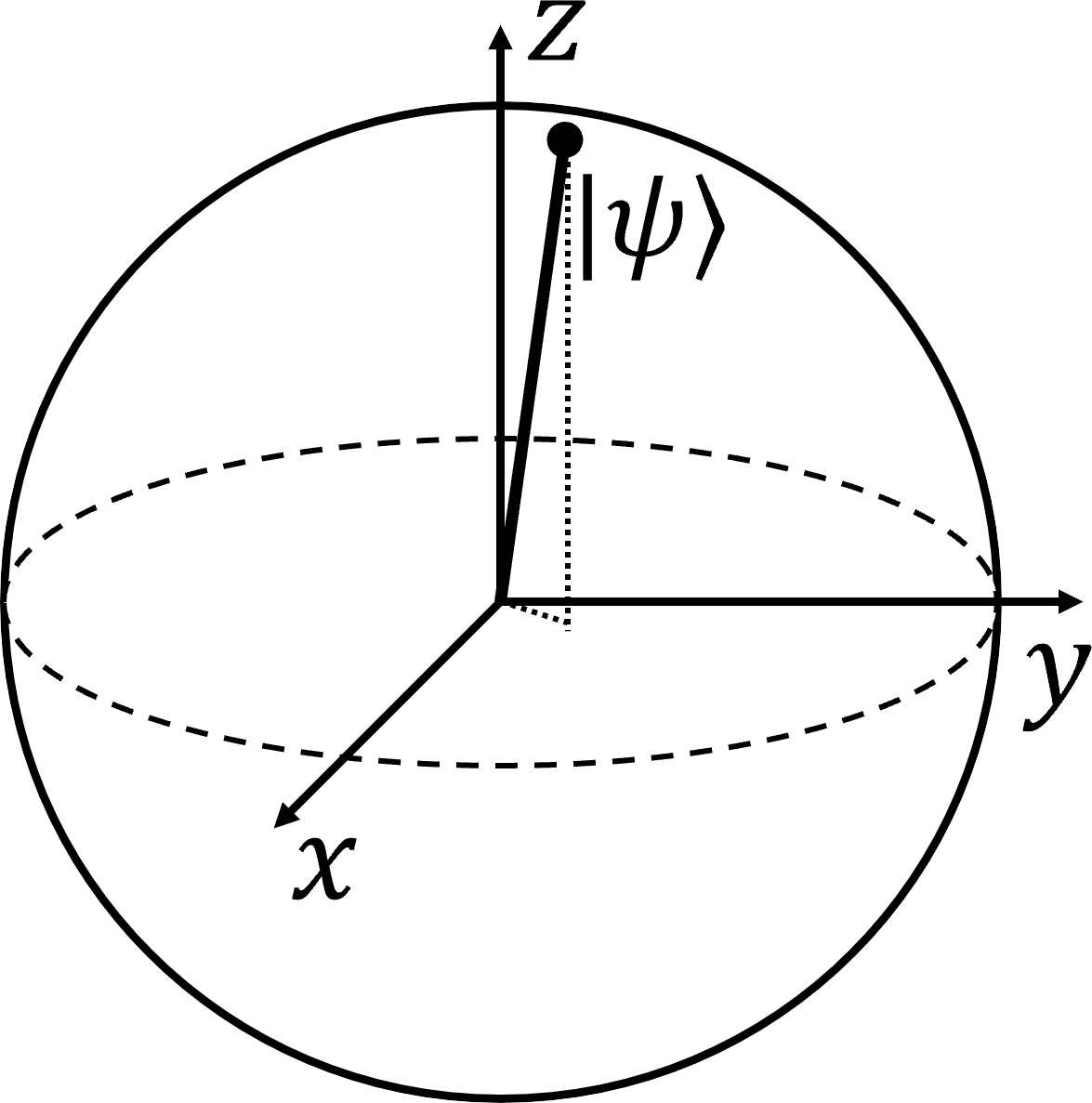} &
\hspace*{-0.15\textwidth}\includegraphics[align=c,width=0.15\textwidth]{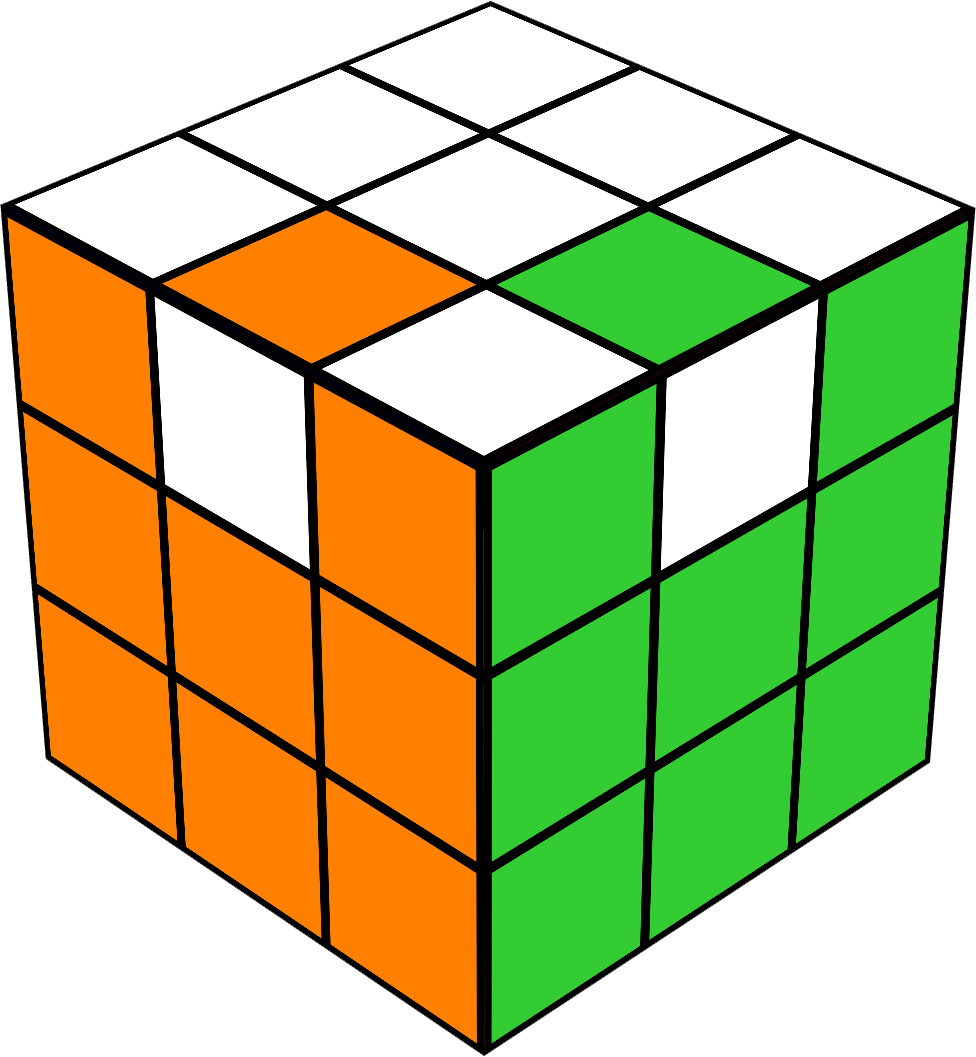}\\\hline
\centered{Initial state} &
\centered{The unitary to be\\ approximated} &
\centered{The scrambled cube} \\\hline
Target state & The identity matrix & The solved cube \\\hline
\centered{Basic move} &
\centered{A gate from the\\ universal set} &
\centered{Rotation of one face}\\
\end{tabular}
\end{ruledtabular}
\caption{Comparison between quantum compiling and solving the Rubik's cube shows that these two seemingly irrelevant problems have a lot in common. The shown cube only has two misplaced cubelets, yet it takes at least 16 steps \cite{rokicki2014diameter} to solve. Similarly, the shown quantum state $|\psi\rangle$ is close to the target state $|0\rangle$, but with a discretized universal gate set, it may still take many steps to transform into $|0\rangle$. \label{tab:comparison}}
\end{table}

Machine learning, especially deep learning, has achieved dramatic success in a broad range of artificial intelligence applications, ranging from image/speech recognition to self-driving cars \cite{Lecun2015Deep,Jordan2015Machine}. The interplay between machine learning and quantum physics has led to an emergent research frontier of quantum machine learning, which has attracted tremendous attention \cite{Biamonte2017Quantum,Sarma2019Machine,Dunjko2018Machine,Carleo2019Machine}. Quantum learning algorithms with potential exponential advantages have been proposed, and machine learning techniques have also been invoked in various applications in quantum physics, including representing quantum many-body states \cite{carleo2017solving,Gao2017anEfficient}, quantum state tomography \cite{Torlai2018Neural, Carrasquilla2019Reconstructing}, non-locality detection \cite{Deng2018MachineNonlocality}, and learning phases of matter \cite{Zhang2016Triangular,Carrasquilla2017Machine,van2017Learning,Wang2016Discovering,Broecker2017Machine,Chng2017Machine,Zhang2017Machine, Wetzel2017Unsupervised,Hu2017Discovering, zhang2019machine,Lian2019Machine}, etc. In this work, we introduce deep reinforcement learning \cite{Sewak2019Deep}, which has been exploited to build AlphaGo \cite{Silver2016AlphaGo} (a computer program of Go that defeated the world's best players) and more recently DeepCubeA that solves the Rubik's cube --- a classic combinatorial puzzle that posed unique challenges for artificial intelligence \cite{agostinelli2019solving} to the task of quantum compiling. We observe that compiling unitaries to a sequence of elementary gates is analogous to finding a sequence of basic moves that solves the Rubik's cube (see Table \ref{tab:comparison}). Since unitaries are invertible, finding a gate sequence approximating a target unitary $U$ is equivalent to finding a gate sequence that ``restores" $U$ back to identity. In this way, the identity matrix becomes our target state (corresponding to the solved cube), and the unitary $U$ is the initial state (corresponding to a scrambled cube). Both problems have several discretized, non-commuting operations; the goal of both problems is to find the shortest sequences available; for a state seemingly close to the targeted one, the actual number of required operations may still be surprisingly large. Similar to the fact that DeepCubeA can solve an arbitrarily scrambled cube in a near-optimal fashion \cite{agostinelli2019solving}, our algorithm can efficiently compile an arbitrary unitary into a near-optimal sequence of elementary gates.

\emph{The algorithm}\textemdash
First, we introduce our general algorithm; later, we will apply it to the case of topological compiling of Fibonacci anyons as a concrete example. In previous reinforcement learning algorithms such as deep Q-learning \cite{mnih2015human, Silver2016AlphaGo}, a function approximator such as a deep neural network (DNN) represents a reward function defined on all states, which dictates the strategy to maximize the reward and performs the actions step-by-step. Then, the resulting experiences are added to the regression to optimize the DNN further, and so on so forth. However, when such an algorithm is directly applied to bring an arbitrary quantum state to a specific target, it faces immediate failure: with a large state space, discretized actions at each step, a single target, and only giving the reward extremely close to the target, the reward may never be received at all, making it almost impossible to train a valid reward function.

To resolve this issue, we start from the target state instead and perform backward search operations, similar to the value iteration algorithm \cite{puterman1978modified}. The cost-to-go function $J(s)$ is defined as the minimum cost for a state $s$ to reach the target state within the designated precision, represented approximately by a DNN. During training, we update the cost-to-go function according to \cite{agostinelli2019solving}:
\begin{equation}
J'(s)=\mathrm{min}_a(g(s, a)+J(S(s, a)))  \label{eq:cost-to-go}
\end{equation}
where $S(s, a)$ is the state obtained after applying the action $a$ to the state $s$, and $g(s, a)$ is the corresponding cost. $J(s_0)=0$ for the target state $s_0$, and $J(s)$ for other states can be computed with Eq. \eqref{eq:cost-to-go} successively. In practice, Eq. \eqref{eq:cost-to-go}
uses the DNN itself for target updating, which may lead to instabilities. Therefore, we use two neural networks during training \cite{mnih2015human, agostinelli2019solving}: a policy network that is constantly being trained, and a target network that estimates of the target value $J'(s)$ for training and only updates to the policy network periodically.

To enhance the search performance and derive the shortest sequence possible, we further complement the cost-to-go function $J(s)$ with a weighted $A*$ search algorithm \cite{hart1968formal, pohl1970heuristic}. We define an evaluation function $f(s)$ from the initial state $s_i$ to the target state $s_0$ via an intermediate state $s$:
\begin{equation}
f(s)=\lambda G(s)+J(s)\label{eq:evaluation}
\end{equation}
where $G(s)$ is the actual cost from the initial state $s_i$ to the current state $s$. $\lambda\in[0, 1]$ is a weighting factor, and smaller $\lambda$ reduces the number of states evaluated and alleviates the difficulty of a large state space at the expense of potentially longer paths \cite{pohl1970heuristic}. During the search, we start with a set of the intermediate states $\{s\}$ with only the initial state $s_i$; iteratively, we pick the state $s$ in $\{s\}$ with the minimum $f(s)$ and replace it with its successors $S(s, a)$ (if they are not already in or have not previously been in $\{s\}$), see Fig. \ref{fig:algorithm schematic}; once a state with a distance less than a designated termination accuracy $\epsilon_T$ from the target state $s_0$ is present in $\{s\}$, we have obtained the desired sequence between $s_i$ and $s_0$ within the desired accuracy threshold.

We also make several additional modifications to the weighted $A*$ search algorithm to better fit our quantum compiling problem. First, we introduce a maximum searching depth $D_{\text{max}}$, beyond which the search terminates and returns the best state found so far. This cut off resolves the possible non-convergence induced by the search along a discrete graph on a continuous state space. Second, it is natural for the DNN to generalize the cost-to-go function $J(s)$ to states never present in training. Sometimes, such a state is mistaken for a small $J(s)$ (e.g., $J(s)\sim 1.5$), although its actual distance from the target state is considerably farther away, and the weighted $A*$ searches are stuck there. To handle this problem, we introduce a decimal-penalty term to the evaluation function $f(s)=\lambda G(s)+J(s)+d(s)$:
\begin{equation}
d(s)=\gamma \frac{\left[J(s)-\mathrm{round}\big(J(s)\big)\right]^2}{J(s)}
\end{equation}
where $\gamma$ is a constant tuning parameter. $d(s)$ put preferences on states used to train the DNN with near-integer $J(s)$ over states whose $J(s)$ values containing decimal parts and are likely estimations and interpolations.

\begin{figure}
\hspace*{-0.48\textwidth}
\includegraphics[width = 0.46\textwidth]{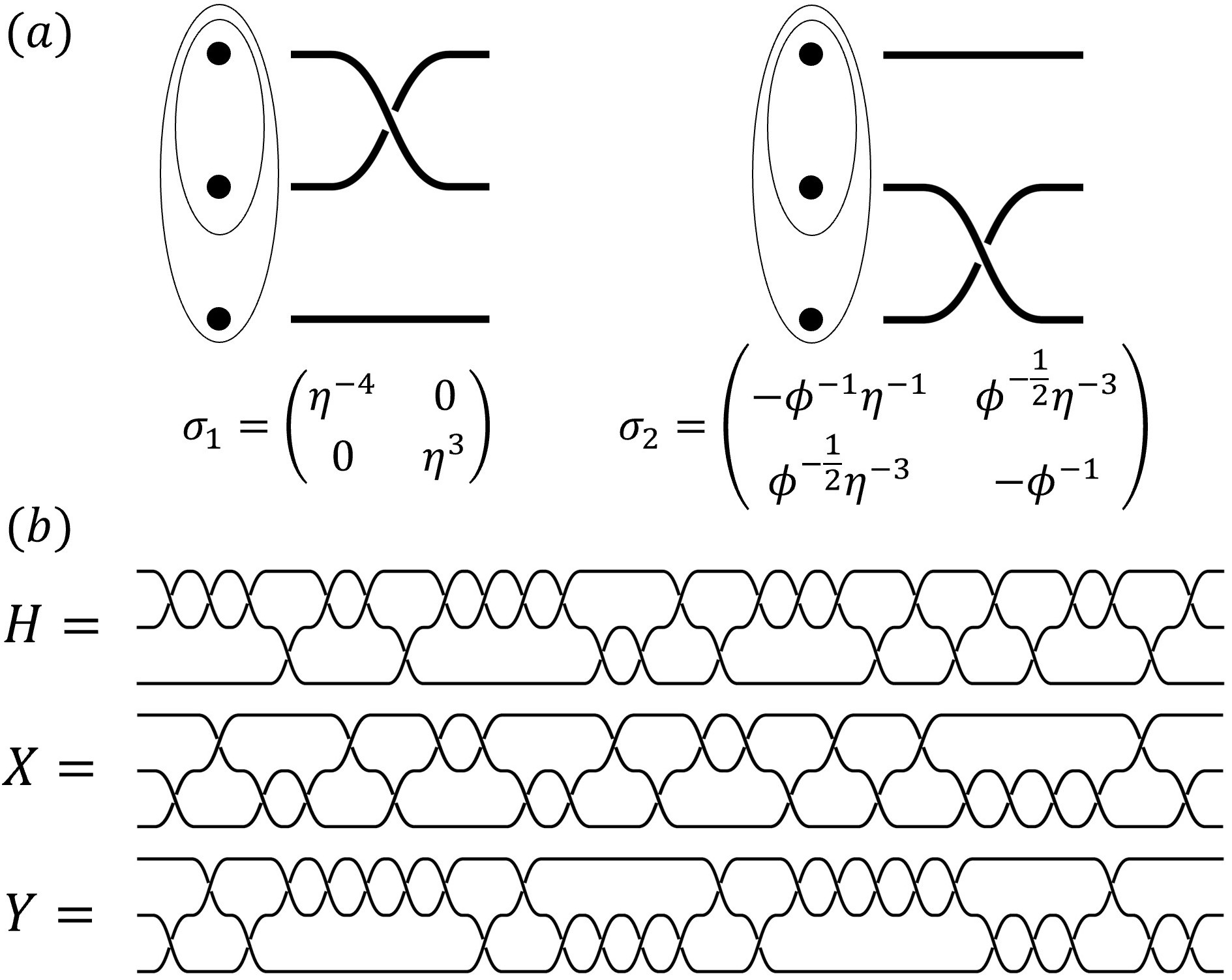}%
\caption{(a) The two elementary gates by braiding Fibonacci anyons. Logical qubits are encoded into triplets of anyons (enclosed in the ovals), and time flows from left to right. Here, $\eta=e^{i\pi /5}$ and $\phi=\frac{\sqrt{5}+1}{2}$ \cite{Chetan2008tqc}. (b) The approximate braiding sequences obtained by the reinforcement-learning based algorithm for the Hadamard gate (H), the Pauli $\sigma^x$ gate (X), and the Pauli $\sigma^y$ gate (Y), respectively. The quaternion distances between the braiding sequences and their corresponding targets are $4.4\times 10^{-3}$, $2.4\times 10^{-3}$, and $2.3\times 10^{-3}$, respectively. After the DNN is trained, the algorithm takes only a couple of seconds to output each of the sequences. }\label{fig:braid_HXY}
\end{figure}

Without loss of generality, we apply our algorithm to topological compiling with Fibonacci anyons, which are quasiparticle excitations of topological states that obey non-Abelian braiding statistics \cite{Chetan2008tqc}. Unlike Majorana bound states \cite{DasSarma2015Majorana}, whose braiding only gives elementary gates in the Clifford group unless additional multi-step protocols are incorporated \cite{Freedman2016PRX}, Fibonacci anyons are the simplest non-Abelian quasiparticles that enable universal topological quantum computation \cite{kitaev2003fault,kitaev2006anyons} by braiding alone \cite{Freedman2002Modular}. They are theoretically predicted to exist in the $\nu=12/5$ fractional quantum Hall liquid \cite{Xia2004Electron} and rotating Bose condensates \cite{Cooper2001Quantum}, as well as quantum spin systems \cite{Freedman2004Aclass,Fendley2005Realizing}. The only nontrivial fusion rule for Fibonacci anyons reads $\tau\times \tau=\mathbf{I} + \tau$, where $\mathbf{I}$ and $\tau$ denote the vacuum and the Fibonacci anyon, respectively. We encode logical qubits into triplets of anyons with total topological charge one  \cite{Freedman2002Modular}: $|0_L\rangle=|((\bullet, \bullet)_{\mathbf{I}}, \bullet)_{\tau}\rangle$ and $|1_L\rangle=|((\bullet, \bullet)_{\tau}, \bullet)_{\tau}\rangle$, and neglect the non-computational state $|NC\rangle=|((\bullet, \bullet)_{\tau}, \bullet)_{\mathbf{I}}\rangle$ since we mainly focus on braidings within a single logical qubit and the leakage error is not relevant in this case. Based on this encoding scheme, the two elementary single-qubit gates correspond to the braidings of two Fibonacci anyons are $\sigma_1$ and $\sigma_2$ as shown in Fig. \ref{fig:braid_HXY}(a), which form a universal set for single-qubit unitaries.

In the literature, topological compiling with Fibonacci anyons has been extensively studied, and different algorithms have been proposed \cite{bonesteel2005braid, xu2008constructing, hormozi2007topological,deng2010fault,burrello2010topological,kliuchnikov2014asymptotically, carnahan2016systematically}. Notable examples include the quantum hashing algorithm \cite{burrello2010topological}, which runs in $O(\log(1/\epsilon))$ time and output a sequence of length $O(\log ^2(1/\epsilon))$, and the probabilistically polynomial algorithm \cite{kliuchnikov2014asymptotically}, which runs in $O(\text{polylog} (1/\epsilon))$ time on average and outputs an asymptotically depth-optimal sequence of length $O(\log (1/\epsilon))$. Here we apply the introduced reinforcement-learning algorithm. To measure the accuracy of the output sequence, we use the quaternion distance \cite{huynh2009metrics}: $d(q_b, q_t) = \sqrt{1 - \langle q_b, q_t\rangle^2}$, where $q_b$ and $q_t$ are the unit quaternions corresponding to the unitary from the braiding sequence and the target unitary respectively, and $\langle q_b,q_t\rangle$ denotes their inner product. We employ a DNN with the state $s$ as the input, two fully-connected hidden layers, and six residue blocks \cite{he2016deep}, followed by one output neuron representing the approximate cost-to-go function $J(s)$. We train this DNN via PyTorch routines with randomly sampled sequences whose lengths are shorter than a given constant \cite{SuppTQCviaRL}.  The training process takes about two days running on a NVIDIA TITAN V GPU. Without loss of generality, we set $g(s, a)=1$ for all gates in Eq. \eqref{eq:cost-to-go}. % into
%\begin{equation}
%J'(s)=\mathrm{min}_a(J(S(s, a)))+1.
%\end{equation}
In situations where certain elementary gates are harder to implement, or the cost is state-dependent, we can simply adjust $g(s, a)$ and retrain the DNN. The optimal values for parameters $\lambda$, $\gamma$ in the evaluation function $f(s)$ and the maximum searching depth $D_{\text{max}}$ are determined by a grid search (see \cite{SuppTQCviaRL}).  Unless noted otherwise, we set $\lambda=1$, $\gamma=400$, and $D_{\text{max}}=100$.

\begin{figure}
\hspace*{-0.48\textwidth}
\includegraphics[width = 0.46\textwidth]{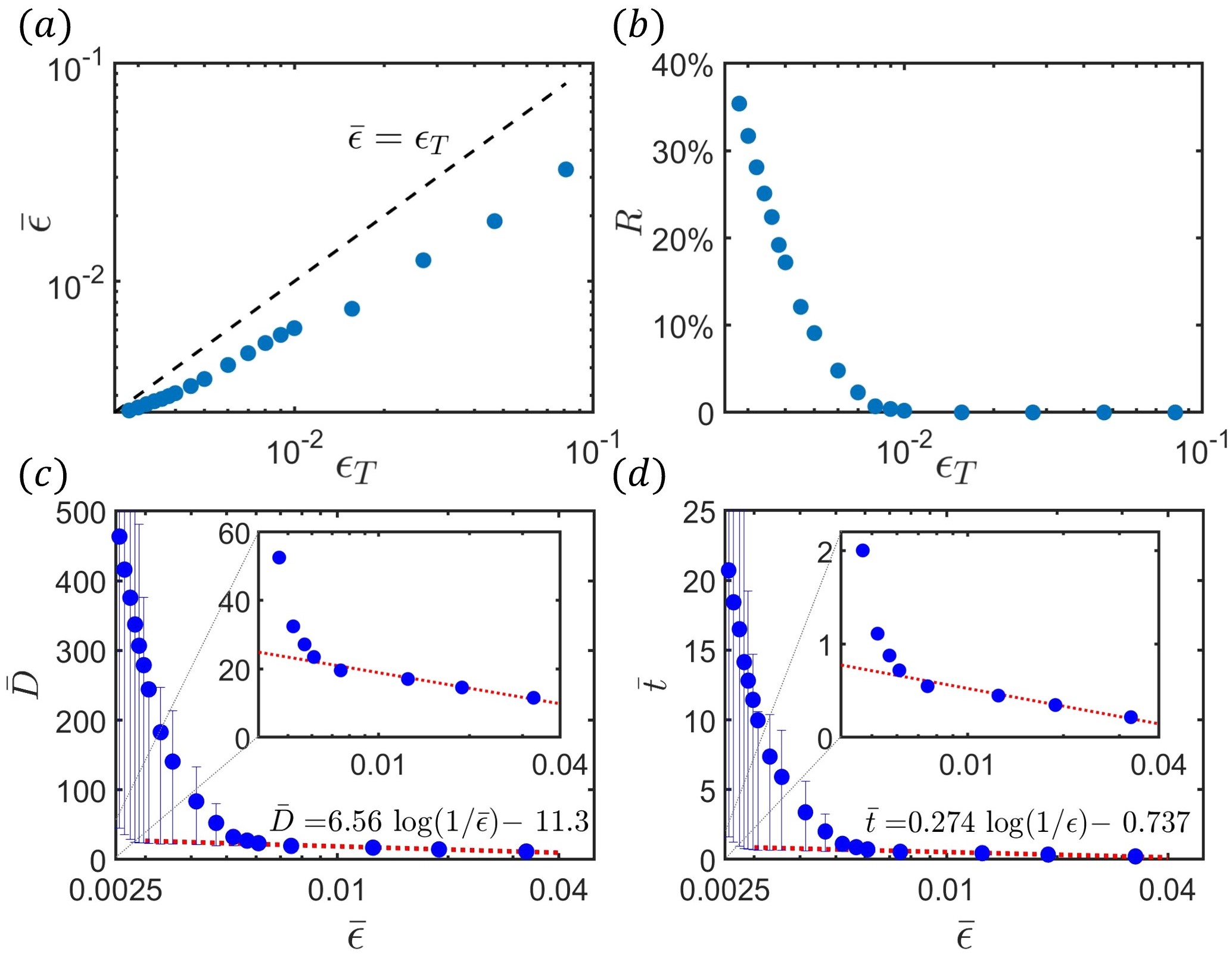}%
\caption{(a) The averaged actual accuracy ($\bar{\epsilon}$) versus the termination accuracy ($\epsilon_T$). Since the distribution of the accuracy has a long tail,  we employ the typical average $\bar{\epsilon}=\exp(\overline{\log\epsilon_i})$. Here, the weighted $A*$ search terminates once a state with an accuracy smaller than $\epsilon_T$ is found, or the searching depth exceeds $D_{\text{max}}=1000$. (b) The ratio ($R$) of target unitaries that require a depth larger than $D_{\text{max}}$ to attain an accuracy smaller than $\epsilon_T$. In general, we need longer sequences for higher accuracy (c) The average searching depth ($\bar{D}$) as a function of the average actual accuracy ($\bar{\epsilon}$). The red dotted line is a logarithmic fitting of $\bar{D}$ into $\log(1/\bar{\epsilon})$. (d) The average searching time $\bar{t}$ versus the average actual accuracy $\bar{\epsilon}$. The red dotted line is a logarithmic fitting of $\bar{t}$ into $\log(1/\bar{\epsilon})$. In these figures, each data point represents an average on the solutions of $1000$ random unitaries.}
\label{fig:complexity}
\end{figure}

In Fig. \ref{fig:braid_HXY}(b), we show the braiding sequences derived by our reinforcement-learning algorithm to approximate the Hadamard gate, the Pauli $\sigma^x$ gate, and the Pauli $\sigma^y$ gate (up to a trivial global phase). Compare to the brute-force searches in previous works \cite{bonesteel2005braid, hormozi2007topological, deng2010fault}, we have achieved comparable accuracy and sequence length, but with much less computational time. Also, we randomly generate $1000$ unitaries in $SU(2)$ and use the reinforcement-learning algorithm to generate their corresponding braiding sequences. The algorithm can efficiently output the desired sequence for any of these unitaries with a running time of less than a couple of seconds on a single GPU. The typical average length of these sequences is  $\sim 24.79$, and the average precision is $\sim 3.1\times10^{-3}$ (see \cite{SuppTQCviaRL}), on par with the results from the brute-force search. To compare our algorithm with the Solovay-Kitaev algorithm, we apply the latter to the same $1000$ unitaries and find the obtained braiding sequences are typically ten times longer.

\begin{figure}
\hspace*{-0.49\textwidth}
\includegraphics[width =0.46\textwidth]{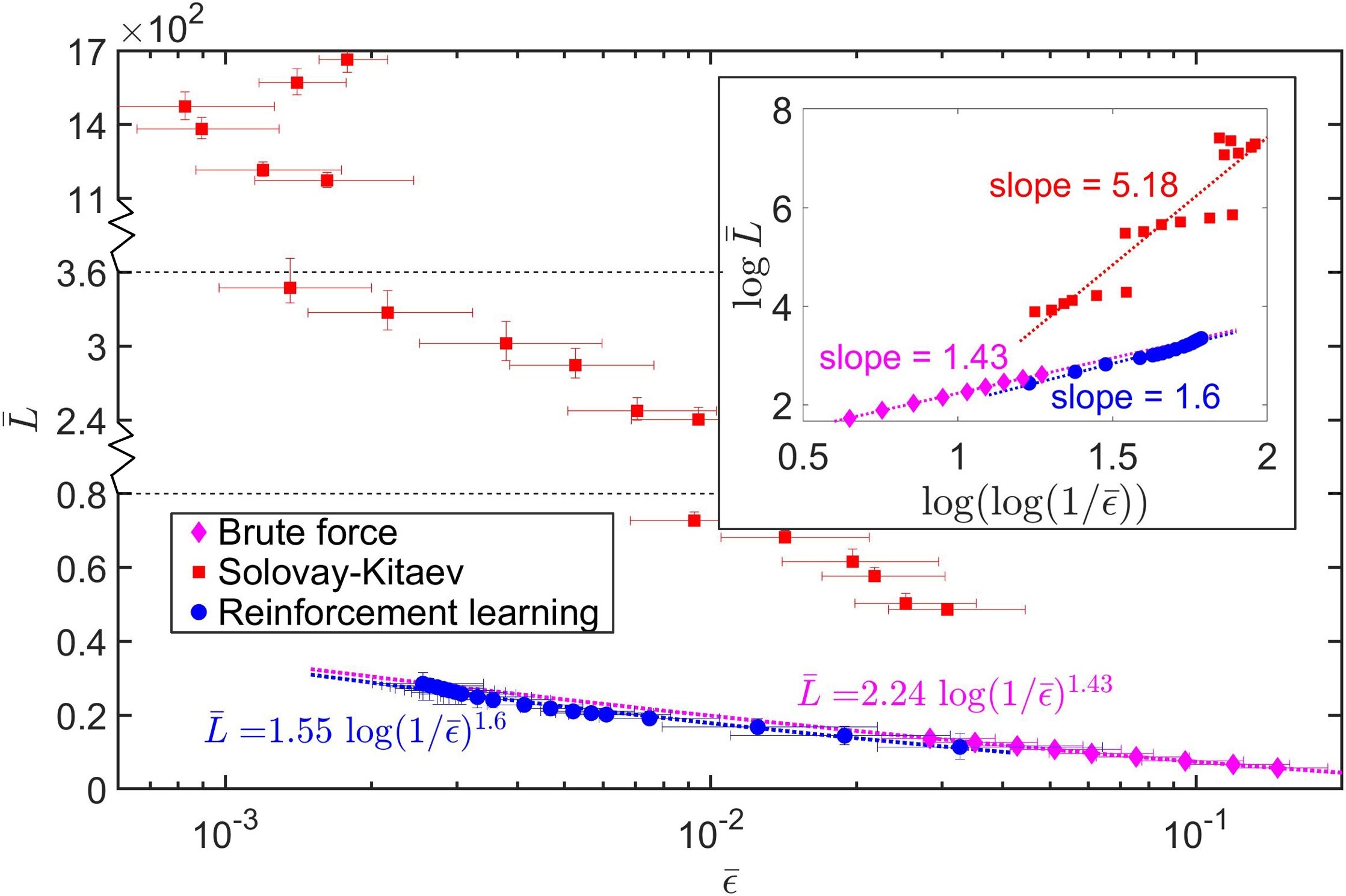}%
\caption{Comparison between different algorithms on the length complexity of the generated gate sequences. Each data point represents an average on the solutions of $1000$ random unitaries, and the error bars show the interval between the lower $25$th percentile and the upper $75$th percentile. Note the breaks in the vertical scale. The inset is a log-log plot for $\bar{L}$ versus $\log(1/\bar{\epsilon})$, and the dotted lines are linear fits of the data.}\label{fig:length complexity}
\end{figure}

To further analyze the time complexity and the length complexity as the scalings of the precision inverse $1/\epsilon$, we explicitly control the approximation accuracy by terminating the weighted $A*$ search once a state with a distance less than $\epsilon_T$ from the target state $s_0$ is found. To ensure that most instances reach the desired accuracy $\epsilon_T$, here we set the maximum searching depth to a larger value $D_{\text{max}}=1000$. Fig. \ref{fig:complexity}(a) shows the averaged actual accuracy $\bar{\epsilon}$ as a function of $\epsilon_T$. When $\epsilon_T$ is large, it is easier to find a sequence with a precision smaller than $\epsilon_T$, and the search terminate before hitting the depth limit $D_{\text{max}}$, thus $\bar{\epsilon}$ is noticeable smaller than $\epsilon_T$. As $\epsilon_T$ becomes smaller, the constraint of limited searching depth becomes dominant, and more and more target unitaries may require the weighted $A*$ searches with a depth larger than the given $D_{\text{max}}=1000$ to attain an accuracy smaller than $\epsilon_T$, as shown in Fig. \ref{fig:complexity}(b). We plot the averaged searching depth $\bar{D}$ as a function of $\bar{\epsilon}$ in Fig. \ref{fig:complexity}(c). From this figure, when $\bar{\epsilon}$ is large, $\bar{D}$ scales logarithmically with $1/\bar{\epsilon}$: $\bar{D}\sim 6.56\log(1/\bar{\epsilon})$, leading to a nearly linear time complexity --- the search time scales as $\bar{t}\sim 0.274\log (1/\bar{\epsilon})$, see Fig. \ref{fig:complexity}(d). As $\bar{\epsilon}$ decreases further, however, the searching depth and time start to increase dramatically. This is likely due to the relatively limited sequence length (no larger than $D\sim 40$) during the training, thus the DNN has not yet learned enough information for approximating unitaries with higher precision. One way to improve the performance of the algorithm for smaller $\bar{\epsilon}$ is to increase $D_{\text{max}}$. Also, we plot the average length $\bar{L}$ of the braiding sequences obtained by different algorithms as a function of $\bar{\epsilon}$ in Fig. \ref{fig:length complexity}. From this figure, $\bar{L}$ scales as $\bar{L}\sim \log ^{1.6} (1/\bar{\epsilon})$ for our reinforcement-learning algorithm, which is slightly worse than the scaling for the brute-force approach but notably better than that for the Solovay-Kitaev algorithm. We note that one may further improve the performance of the reinforcement-learning algorithm in the above example, through increasing the size of the DNN, the length of the braiding sequences in the training set, or the searching depth when generating sequences, etc. In fact, we used a much smaller DNN in this work than that for AlphaGo \cite{Silver2016AlphaGo} and only a single GPU.

The reinforcement learning algorithm can also compile two- or multi-qubit gates, with enlarged state space (target unitary matrices) and action space (gates in the universal set) accordingly, which demands a larger DNN, and inevitably, increases the cost for its training. For simplicity, here we consider the compiling of arbitrary two-qubit gates for demonstration. The action space involves braiding six Fibonacci anyons within the 87-dimensional Hilbert space, much larger than the case for single-qubit gates \cite{bonesteel2005braid}. Alternatively, we can decompose an arbitrary two-qubit gate into seven single-qubit gates and three controlled-NOT (CNOT) gates analytically and optimally \cite{vatan2004optimal}. In turn, the CNOT gate can be decomposed into a single-qubit rotation and a controlled-iX gate, whose braiding sequence is available \cite{bonesteel2005braid, hormozi2007topological}. Finally, our reinforcement learning algorithm can compile the component single-qubit unitaries. Indeed, this build-up decomposes arbitrary two-qubit gates into braiding sequences with notably better performance than the Solovay-Kitaev algorithm \cite{SuppTQCviaRL}.

\emph{Discussion and conclusion}\textemdash In experiments, it is common that elementary gates cost differently, and reducing the use of the expensive ones in compiling is of crucial importance for applications in quantum computing. Notably, each elementary gate's cost can be naturally incorporated into our reinforcement-learning approach by adjusting the cost function [$g(s, a)$ in Eq. (\ref{eq:cost-to-go})] --- another striking advantage of the proposed approach over traditional algorithms. Moreover, our approach carries over straightforwardly to other quantum control problems \cite{Dong2010Quantum} as well.

In summary, we have introduced a reinforcement-learning-based quantum compiling algorithm to decompose an arbitrary unitary into a sequence of elementary gates from a finite universal set. This algorithm uses no ancillary qubit or group-theory relevance and is generally applicable to various scenarios regardless of the choice of universal gate sets. It generates near-optimal gate sequences that approximate arbitrary unitaries to given accuracy in an efficient fashion.  To illustrate how the algorithm works, we have also applied it to topological compiling with Fibonacci anyons. Our results build a new connection between reinforcement learning and quantum compiling, which would benefit future studies in both areas.

\emph{Acknowledgment}\textemdash We acknowledge insightful discussions with Lei Wang. Y.Z. and P.L.Z. are supported by the start-up grant from Peking University. Y.H.Z. and D.L.D. acknowledge the start-up fund from Tsinghua University (Grant. No. 53330300320). DLD also would like to acknowledge additional support from the Shanghai Qi Zhi Institute. The source code for this work can be found at \href{https://github.com/yuanhangzhang98/ml_quantum_compiling}{https://github.com/yuanhangzhang98/ml$\_$quantum$\_$compiling}.

\bibliographystyle{apsrev4-1-title}
\bibliography{bibliography,Dengbib}

\newpage

\appendix
\section{Supplementary Material}
In this Supplementary Material, we provide more details for our algorithm introduced in the main text and more numerical data about applying this algorithm to compiling Fibonacci anyons.

\emph{Neural network architecture.}\textemdash
Our deep neural network (DNN) consists of two hidden layers, six residual blocks \cite{he2016deep}, and one output neuron. The first two hidden layers are of sizes 5000 and 1000, respectively, and each residual block consists of two hidden layers with 1000 hidden neurons each. The activation function is leaky ReLU \cite{maas2013rectifier} throughout the neural network, and batch normalization \cite{ioffe2015batch} is applied in all layers.

\emph{Training the DNN.}\textemdash
The DNN is trained to approximate the cost-to-go function  $J(s)$ of the input unitary matrix. During training, the DNN only utilized the following knowledge \cite{agostinelli2019solving}:
\begin{enumerate}
    \item $J(s_0)=0$ for the identity matrix $s_0$.
    \item For an arbitrary state $s$ other than $s_0$,
    \begin{equation}J(s)=\mathrm{min}_a(J(S(s, a)))+1\label{eq:cost-to-go-S}
    \end{equation}
    where $S(s, a)$ is the state obtained after applying the action $a$ to the state $s$. In other words, the cost-to-go for each state equals to the cost-to-go of its predecessor plus the cost for each action.
\end{enumerate}

Starting from the identity matrix $s_0$ and iterate with the two rules above, the cost-to-go function for every matrix in the state space is uniquely determined. In practice, we set $J(s)=0$ for all the $s$ with $d(s, s_0)<10^{-4}$, where $d(a, b)$ is the quaternion distance \cite{huynh2009metrics} defined in the main text.

Before training, the DNN is initialized with random small weights, and its initial guess for $J(s)$ will be random small numbers. To train the DNN, we feed matrices generated by random gate sequences into the DNN, and optimize the DNN weights by minimizing the loss function $\sum_s[J(s)-\mathrm{min}_a(J(S(s, a))+1)]^2$ with the standard back-propagation algorithm.

To ensure that the DNN gradually explores the state space, we generate the training data by applying $k$ random operations to the identity matrix, where $k$ is evenly distributed between $1$ and a maximum length $M$. We set $M=5$ in the beginning, and increases $M$ by $1$ whenever the loss falls below a pre-defined threshold $\delta$.

Training a neural network with itself as its target causes instabilities. In the machine learning community, the standard technique to overcome this issue is to use two neural networks during training \cite{agostinelli2019solving}: a policy network and a target network. These two neural networks are almost identical, except that only the policy network is being trained with back-propagation. The target network stays intact most of the time and provides estimations of the updated cost-to-go function in Eq. \eqref{eq:cost-to-go-S}. Whenever loss falls below $\delta$, the target network will be updated by copying parameters from the policy network.

The training algorithm is summarized in Algorithm \ref{alg:training}.

\begin{algorithm}[b]
\caption{Training the DNN}
\label{alg:training}
\begin{algorithmic}
\ENSURE The trained DNN parameters $\theta_p$
\STATE Initialize the policynet parameters $\theta_p$
\STATE Targetnet parameters $\theta_t \leftarrow \theta_p$
\FOR{$i=1$ to max$\_$epoch}
	\STATE Generate training data $s$
	\STATE $\mathrm{loss} \leftarrow \sum_s\Big(J_p(s)-\mathrm{min}_a(J_t(S(s, a))+1)\Big)^2$
	\STATE loss.backward()
	\STATE optimizer.step()
	\IF{loss$<\delta$}
		\STATE $\theta_t \leftarrow \theta_p$
		\STATE Maximum length $M \leftarrow M+1$
	\ENDIF
\ENDFOR
\end{algorithmic}
\end{algorithm}

\begin{figure}
\hspace*{-0.46\textwidth}
\includegraphics[width = 0.460\textwidth]{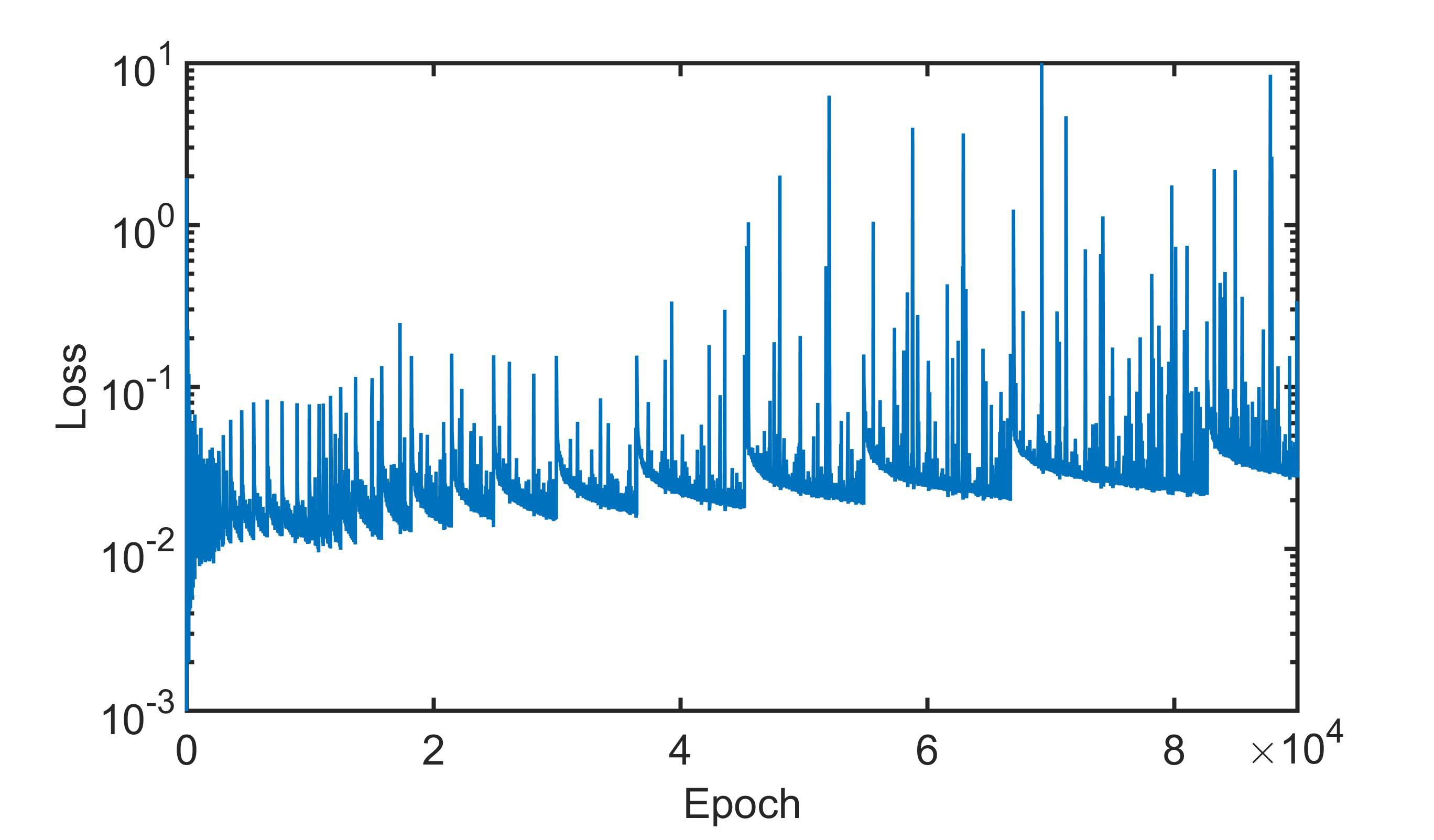}%
\caption{The loss curve during training. The loss function spikes up whenever we raise the maximum length $M$, then training gradually brings the loss back until it falls below the threshold $\delta$, after which we raise $M$ again. Here and throughout this whole Supplementary Material, the data plotted are all based on topological compiling with Fibonacci anyons.   \label{fig:loss}}
\end{figure}

\begin{figure}
\hspace*{-0.40\textwidth}
\includegraphics[width = 0.40\textwidth]{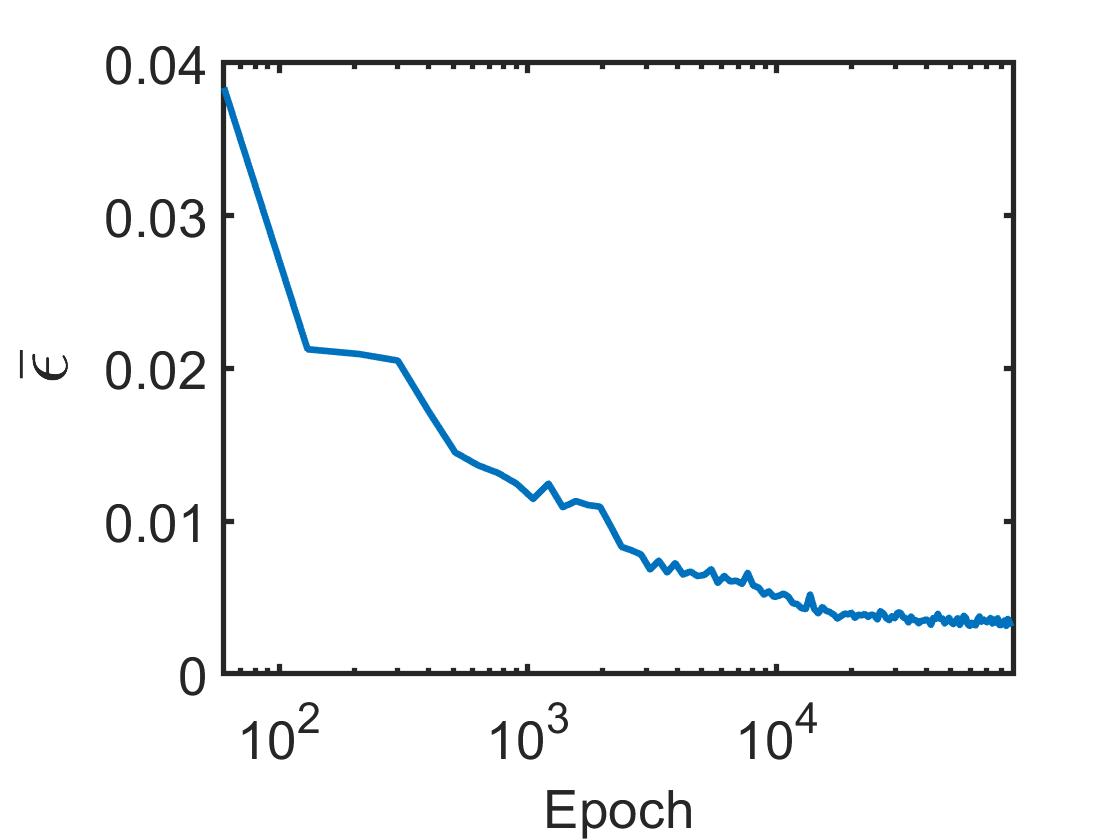}%
\caption{The performance of the DNN during different training stages, evaluated as the averaged approximation accuracy $\bar{\epsilon}$ of gate sequence solutions for 100 randomly generated SU(2) unitaries. The searching depth $D_\text{max}=100$. \label{fig:performance_curve}}
\end{figure}

Fig. \ref{fig:loss} shows the loss curve during training. For reinforcement learning, large fluctuations in the loss function are common: whenever the DNN sees new data or updates the target network, the loss spikes up. Nevertheless, the performance of the DNN is stable. Fig. \ref{fig:performance_curve} shows the average accuracy $\bar{\epsilon}$ of the gate sequences generated by our DNN at different training stages, in which $D_\text{max}=100$ and each data point is calculated by averaging loss over 100 sample instances. $\bar{\epsilon}$ saturated after $2\times 10^4$ epochs of training, and did not improve much afterwards. We trained this model on a single NVIDIA TITAN V GPU for about a week, but the result indicates that two days' training is already sufficient.

\begin{algorithm}[b]
\caption{Generating gate sequence with $A*$ search}
\label{alg:searching}
\begin{algorithmic}
\REQUIRE The target gate to be approximated, $T$
\ENSURE A gate sequence $U_1U_2\cdots U_n\approx T$
\STATE $\{s\}$ $\leftarrow \{T\}$
\FOR{$i=1$ to $D_{\text{bf}}$}
	\STATE $\{s\}$ $\leftarrow$ the universal gate set $\{U_i\}\times$ $\{s\}$
	\STATE ($\times$ denotes Cartesian product)
	\STATE Update and store $s_{\text{best}}$, the matrix closest to identity
\ENDFOR
\FOR{$i=1$ to $D_{\text{max}}$}
	\STATE Choose $N$ states $\{S\}$ with minimum $f(s)$ from $\{s\}$
	\STATE $\{S'\} \leftarrow \{U_i\}\times\{S\}$
	\STATE Update $s_{\text{best}}$
	\STATE Add $\{S'\}$ to $\{s\}$ and remove $\{S\}$
	\IF{$|\mathrm{\{s\}}|>\mathrm{maximum}\_\mathrm{size}$}
		\STATE Remove the states with maximum $f(s)$
	\ENDIF
\ENDFOR
\STATE Return $s_{\text{best}}$ and its corresponding gate sequence
\STATE Inverse the gate sequence
\end{algorithmic}
\end{algorithm}

\emph{Training data generation.}\textemdash
To train the DNN, we need unitary matrices generated by random gate sequences, but a vast amount of data is required. To generate gate sequences more efficiently, we first performed a brute-force expansion, evaluating all possible gate sequences to a predefined brute-force depth. After that, further operations are generated randomly, until the length of the gate sequence reaches the predefined length $k$. During this procedure, we store all the encountered matrices in the memory for later training, and the brute-force expansion depth is adjusted according to available memories.

We mention that the generation of training data in this way is very efficient since its complexity is mainly about two-by-two matrix multiplications. In fact, the whole data generation procedure is carried out in parallel on a single GPU, and takes only less than one second. The matrices generated by random sequences are periodically updated during training.

\emph{Weighted $A*$ search with a decimal penalty term.}\textemdash
$A*$ search \cite{hart1968formal} is a heuristic searching algorithm that finds a path between an initial node $s_i$ and a target node $s_0$ on a graph. The evaluation function that $A*$ search employs is defined as $f(s)=G(s)+J(s)$, where $G(s)$ is the path cost (the cost already taken to reach $s$ from $s_i$), and $J(s)$ is a heuristic function that estimates the cost still required to reach $s_0$ from $s$, or the cost-to-go function. During the search, the $A*$ algorithm maintains a set of intermediate states $\{s\}$. The algorithm starts with only $s_i$ in $\{s\}$. In each iteration, the algorithm selects and removes from $\{s\}$ the $s$ with the smallest $f(s)$, and adds in its successors if they have never been in $\{s\}$ previously. This procedure repeats until the target node $s_0$ appears in $\{s\}$.

\begin{figure}
\hspace*{-0.43\textwidth}
\includegraphics[width = 0.43\textwidth]{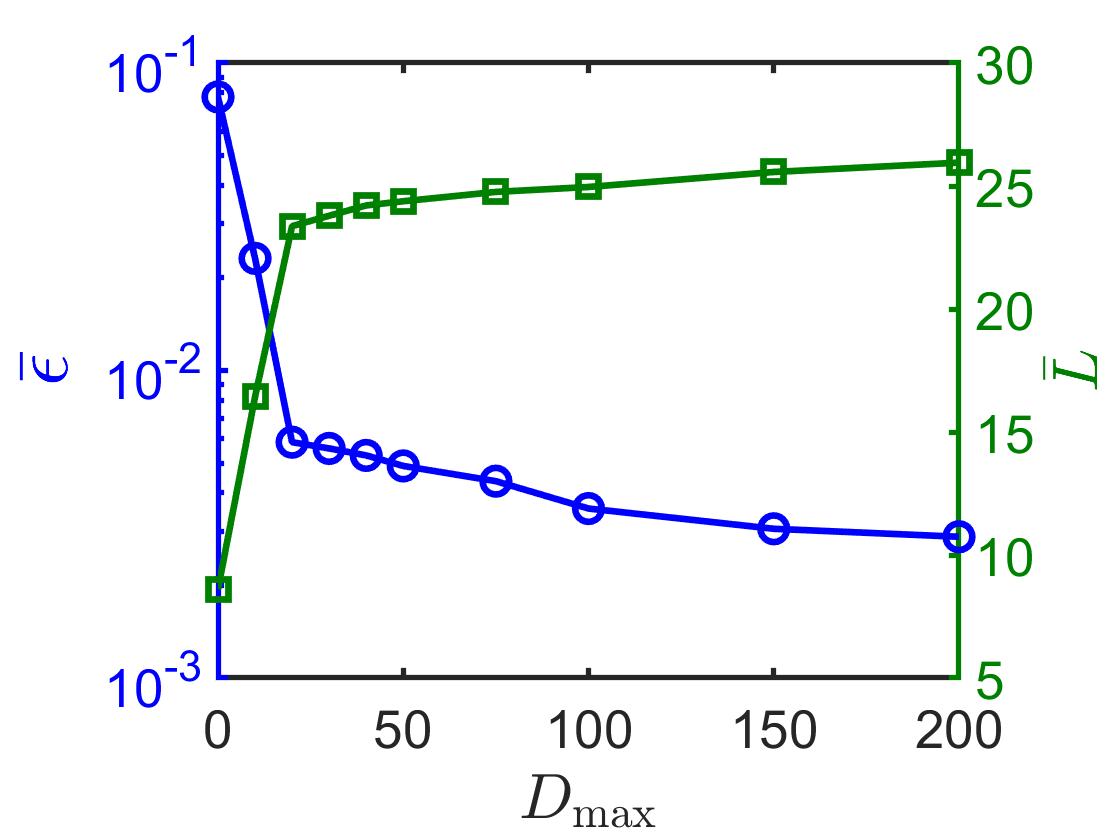}%
\caption{We search for the optimal searching depth $D_\text{max}$ by plotting the average accuracy $\bar{\epsilon}$ and the average braiding sequence length $\bar{L}$ against $D_\text{max}$, and choose $D_\text{max}=100$ to balance between approximation accuracy and time cost. Higher accuracy can be achieved by increasing $D_\text{max}$ at the cost of computation time. \label{fig:maximum_depth}}
\end{figure}

It is proved that $A*$ search can find the shortest possible path if $J(s)$ never overestimates the actual cost \cite{hart1968formal}. However, as $A*$ search stores all the nodes it generates, in general, it has an exponential space complexity, making it unpractical on large graphs. Weighted $A*$ search \cite{pohl1970heuristic} partially resolves this issue by defining the evaluation function as $f(s)=\lambda G(s) + J(s)$, where $\lambda$ is a weighting factor between 0 and 1. This method trades potentially longer paths for potentially less memory usage, alleviating the memory burden on large graphs.

For our problem of quantum compiling, we have an infinite graph in a continuous state space, and $A*$ search may face convergence issues. To resolve this, we imposed a maximum searching depth $D_{\text{max}}$: searching terminates after reaching $D_{\text{max}}$ and the state closest to the target state is returned. Since the DNN heuristic $J(s)$ can be non-reliable in extreme cases, as explained in the main test, we added an additional decimal penalty term. The evaluation function reads:
\begin{equation}
f(s)=\lambda G(s)+J(s)+\gamma \frac{\Big(J(s)-\mathrm{round}\big(J(s)\big)\Big)^2}{J(s)} \label{eq:eval}
\end{equation}

In order to better utilize the parallelization of GPU, we add a short brute-force search phase in the beginning, starting with the initial matrix and apply all possible gate sequences onto it up to a length of $D_{\text{bf}}$, storing all encountered states. Using the states generated by brute-force search as the initial set of intermediate states $\{s\}$, we evaluate their $f(s)$ in parallel and expand $N$ states with the smallest $f(s)$, removing them from $\{s\}$ while adding their successors in, and so on so forth. Whenever the size of $\{s\}$ reaches a maximum value, the states with the largest $f(s)$ are discarded. The procedure terminates after the number of searching steps reaches the maximum searching depth $D_{\text{max}}$, upon which the gate sequence closest to the target state is returned. The searching algorithm is summarized in Algorithm \ref{alg:searching}. In the main text, we also have our algorithm terminate once a state with accuracy $\epsilon < \epsilon_{T}$ is found. Together with a larger $D_{\text{max}}$ that is no longer a limiting factor, we can reach the desired accuracy and analyze the time and length complexity of our algorithm.

\begin{figure}
\hspace*{-0.42\textwidth}
\includegraphics[width = 0.42\textwidth]{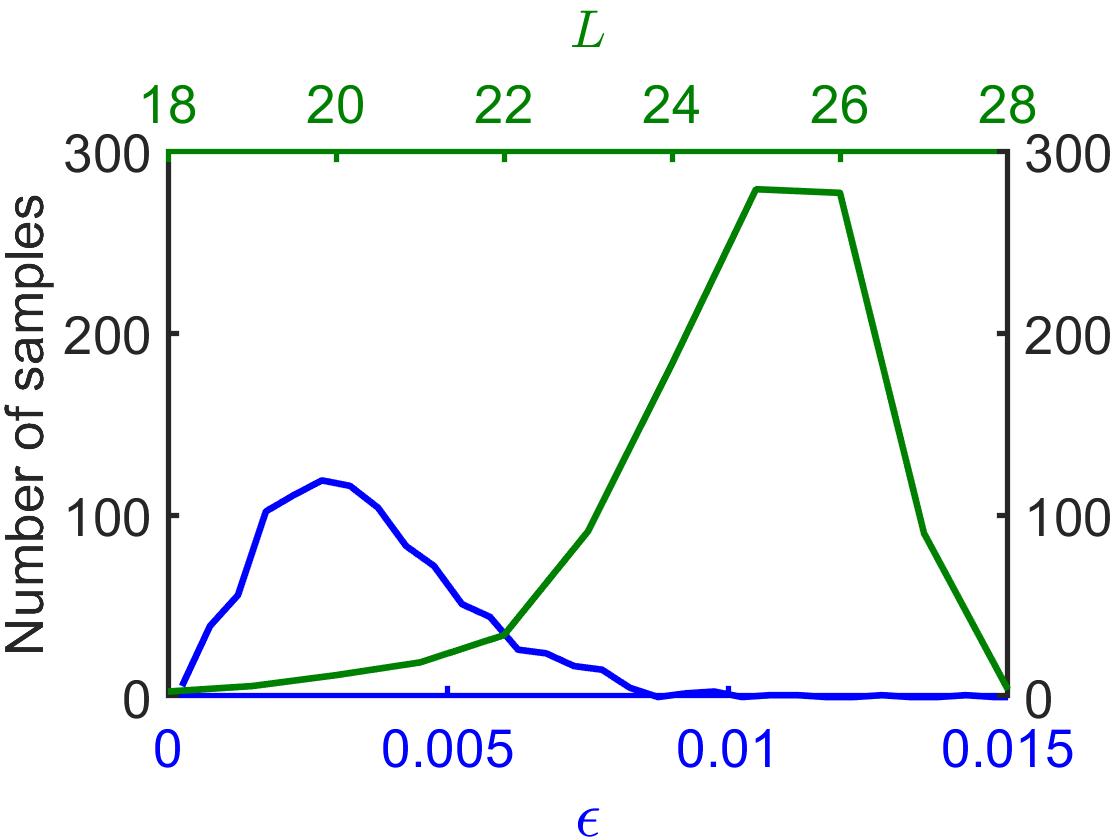}%
\caption{The distribution of the approximation accuracy $\epsilon$ and the sequence length $L$ on the solution of 1000 randomly generated $SU(2)$ matrices using our reinforcement learning algorithm, where the searching depth $D_{\text{max}}=100$. Note that the distribution of $\epsilon$ has a long tail, while the distribution of $L$ is more close to a Gaussian distribution. To better characterize $\bar{\epsilon}$, we adopt the typical average $\bar{\epsilon}=\exp(\overline{\log\epsilon_i})$. In this figure, $\bar{\epsilon}=0.0031$, and $\bar{L}=24.79$. \label{fig:dist_len}}
\end{figure}

%\section{More data}

\begin{figure}[t]
\hspace*{-0.47\textwidth}
\includegraphics[width = 0.47\textwidth]{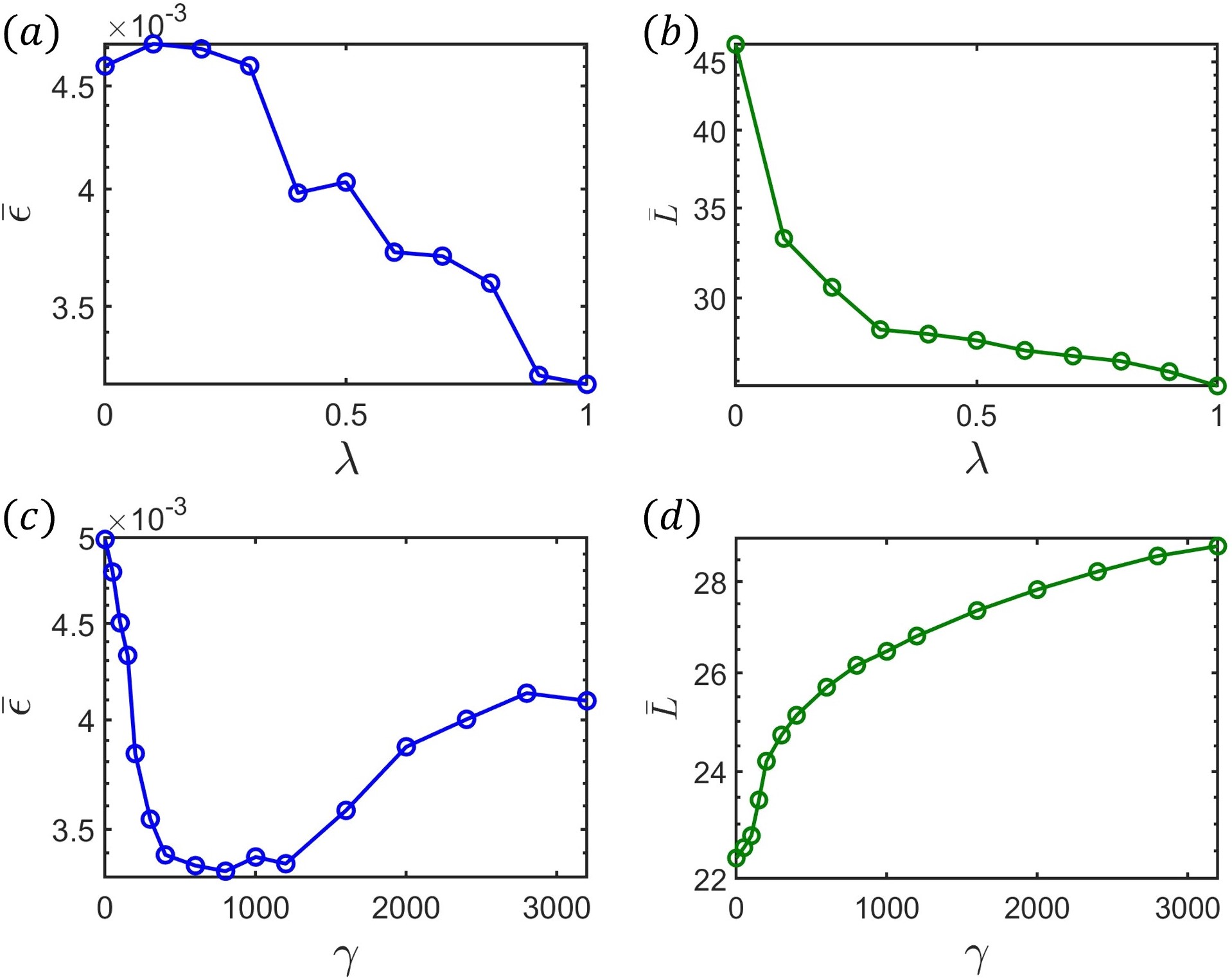}%
\caption{$(a)$ The average accuracy $\bar{\epsilon}$ and $(b)$ average sequence lengths $\bar{L}$ versus $\lambda$ for $\gamma=400$. $\lambda=1$ leads to the shortest sequence and the highest accuracy. $(c)$ The average accuracy $\bar{\epsilon}$ and $(d)$ average sequence lengths $\bar{L}$ versus $\gamma$ for $\lambda=1$. The accuracy peaked around $\gamma\sim 1000$, while the sequence length increases monotonically with $\gamma$. \label{fig:decimal_lambda}}
\end{figure}

The computation time required for each searching step is roughly a constant, and $D_{\text{bf}}$ is small so that the time for the initial brute-force search is negligible. Therefore, $D_{\text{max}}$ is directly proportional to the time cost and responsible for the time complexity of our algorithm. Fig. \ref{fig:maximum_depth} shows how the average accuracy $\bar{\epsilon}$ and sequence length $\bar{L}$ change with $D_{\text{max}}$. We can see that the accuracy improves dramatically at $D_{\text{max}} < 30$ and starts to saturate after $D_{\text{max}}\sim 100$, while the sequence length also reaches a plateau after $D_{\text{max}}\sim 50$. Considering time efficiency, we set $D_{\text{max}}=100$, which roughly takes three seconds on a single GPU. Higher accuracy can be achieved by further increasing $D_{\text{max}}$ at the cost of computation time.

Fig. \ref{fig:dist_len} shows the distributions of the approximation accuracy $\epsilon$ and the sequence length $L$ on the solution of 1000 randomly generated unitaries with $D_\text{max}=100$. Note that the distribution of $\epsilon$ has a long tail. Therefore, to better characterize $\bar{\epsilon}$, we adopt the typical average $\bar{\epsilon}=\exp(\overline{\log\epsilon_i})$, which is $\bar{\epsilon}=0.0031$, and $\bar{L}=24.79$ in this figure.

% \begin{figure}
% \hspace*{-0.40\textwidth}
% \includegraphics[width = 0.40\textwidth]{lambda_search.jpg}%
% \caption{The average accuracy $\bar{\epsilon}$ and average sequence lengths $\bar{L}$ versus $\lambda$ for $\gamma=400$. $\lambda=1$ leads to the shortest sequence and the highest accuracy.\label{fig:lambda}}
% \end{figure}

The parameters $\lambda$ and $\gamma$ in $f(s)$ are determined with a grid search. In weighted $A*$ search, $\lambda<1$ is a compromise between insufficient memory and optimal paths. Since we manually imposed a maximum searching depth $D_{\text{max}}$, memory is no longer a limiting issue for us, and $\lambda=1$ should yield the best results. Indeed, Fig. \ref{fig:decimal_lambda} $(a)(b)$ verifies that $\lambda=1$ leads to the shortest sequence and the highest accuracy. Note that for longer searching depth $D_{\text{max}}$ and more complex scenarios, the optimal $\lambda$ may become different. Fig. \ref{fig:decimal_lambda} $(c)(d)$ shows the average accuracy $\bar{\epsilon}$ and the average sequence length $\bar{L}$ versus $\gamma$ for $\lambda=1$. The average accuracy first decreases and then increases as $\gamma$ increases,  while the sequence length increases monotonically with $\gamma$. We set $\gamma=400$ to balance between accuracy and sequence length.

\begin{figure}[!htbp]
\hspace*{-0.47\textwidth}
\includegraphics[width = 0.47\textwidth]{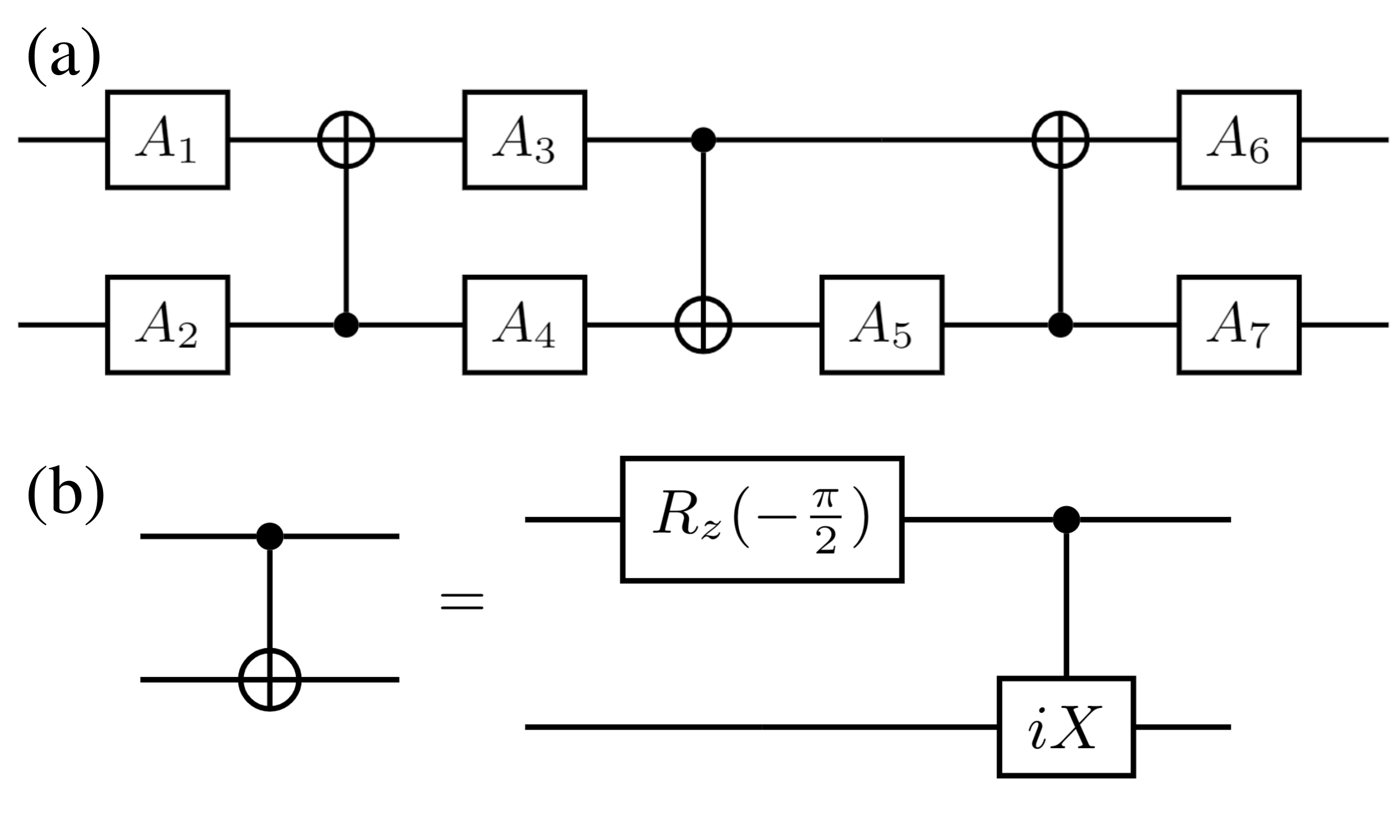}%
\caption{(a) The quantum circuit can decompose an arbitrary two-qubit gate in $SU(4)$ into seven one-qubit gates and three CNOT gates, up to a global phase \cite{vatan2004optimal}. The details of the implementation are available in \cite{vatan2004optimal, kraus2001optimal, tucci2005introduction}. (b) A CNOT gate can be decomposed into a controlled-$iX$ gate and a single-qubit rotation $R_z(-\frac{\pi}{2})$. By replacing the three CNOT gates with $R_z(-\frac{\pi}{2})$ + controlled-$iX$ gate \cite{bonesteel2005braid}, merging the $R_z(-\frac{\pi}{2})$ with the neighboring single-qubit unitaries and compiling them with our reinforcement learning algorithm, we reach our circuit for compiling an arbitrary two-qubit gate, in which every gate can be implemented by braiding Fibonacci anyons.  \label{fig:2qubit_decompose}}
\end{figure}

\emph{Compiling two-qubit gates.}\textemdash
As a second example, we demonstrate how to decompose an arbitrary two-qubit gate into braidings of Fibonacci anyons with our algorithm.

To encode two qubits into Fibonacci anyons, we follow the conventions in \cite{bonesteel2005braid} and represent them with two triplets, or six Fibonacci anyons. In principle, we can repeat our approach for single-qubit gates and train a more powerful DNN with an enlarged action space to compile an arbitrary two-qubit gate. However, the 87-dimensional Hilbert space for six Fibonacci anyons \cite{bonesteel2005braid} is much larger than the three-dimensional Hilbert space for three anyons, which necessarily increases the training cost for the DNN. Unlike the single-qubit unitaries, we are unable to achieve a satisfactory accuracy for two-qubit unitaries with a single GPU within several days.

As an alternative, we can decompose every two-qubit gate into three CNOTs and seven single-qubit gates (Fig. \ref{fig:2qubit_decompose}a) \cite{vatan2004optimal}. This decomposition is analytical and proved optimal within its setup. In the meantime, the braiding sequence for the controlled-$iX$ gate is available \cite{bonesteel2005braid, hormozi2007topological}, which, when combined with a single-qubit rotation, offers the CNOT gate (Fig. \ref{fig:2qubit_decompose}b). The remaining single-qubit unitary components can be compiled using our reinforcement learning algorithm. Altogether, our algorithm can decompose an arbitrary two-qubit gate into braidings of Fibonacci anyons. Note that we have restricted particular gate sequences in the search by this construction scheme, which may impact the overall approximation accuracy and sequence length performance. However, as the degrees of freedom in multi-qubit gates scale exponentially, it may be advisable to include some extent of restriction and analytical decomposition on top of the more generic compiling by the reinforcement learning algorithm at lower level to balance the overall cost and performance. As we will see next, such a joint endeavor already makes a step forward in reducing the computational complexity in the two-qubit gate scenarios.

\begin{figure}[!htbp]
\hspace*{-0.4\textwidth}
\includegraphics[width = 0.4\textwidth]{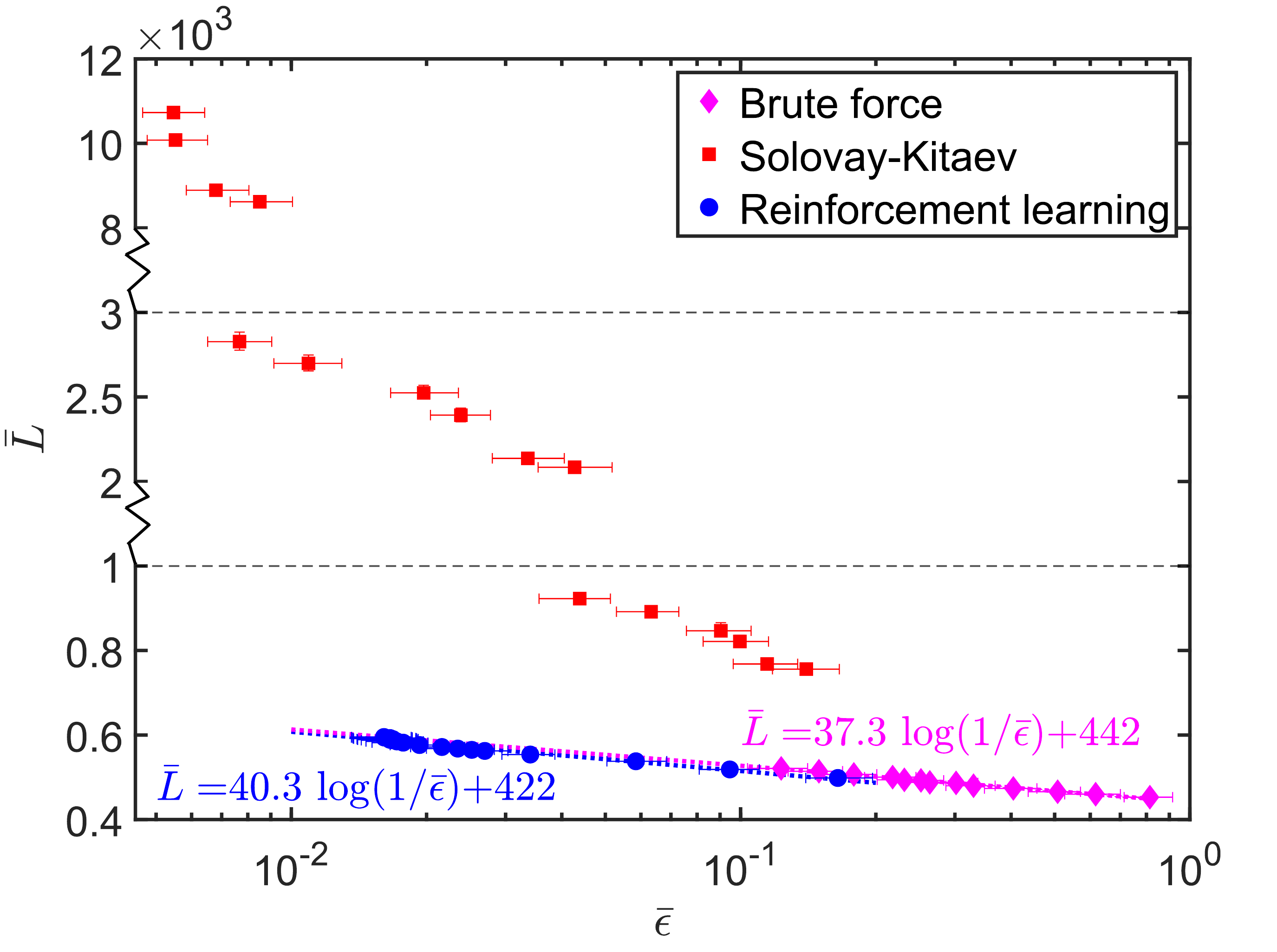}%
\caption{Comparison between different algorithms on the length complexity of the generated braiding sequences when compiling two-qubit gates. All three algorithms use the decomposition in Fig. \ref{fig:2qubit_decompose} and the controlled-$iX$ gate constructed in \cite{bonesteel2005braid}, and the searching algorithms are responsible for compiling the seven component single-qubit gates resulted from the decomposition. Each data point represents an average of $2000$ trials ($1000$ each for the total charge of the braiding anyons being $\mathbf{I}$ or $\tau$) on randomly generated $SU(4)$ matrices, and the error bars show the interval between the lower 25th and upper 75th percentile. Note the breaks in the vertical scale. Unlike Fig. 4 in the main text, we didn't perform the power-law fit, as the length and error of the approximated controlled-$iX$ gate interrupts the power-law scaling. Instead, we show that the sequence lengths of both the reinforcement learning algorithm and brute force search admit a linear fit, with reinforcement learning continuing the scaling for the brute force search.  \label{fig:2qubit_acc_comparison}}
\end{figure}

In Fig. \ref{fig:2qubit_acc_comparison}, we compare the performance of the brute force search, the Solovay-Kitaev algorithm, and our reinforcement learning algorithm on compiling two-qubit gates. All three algorithms use the decomposition in Fig. \ref{fig:2qubit_decompose}, and the search algorithms are only responsible for the seven component single-qubit gates resulted from the decomposition. The controlled-$iX$ gate constructed in \cite{bonesteel2005braid} has a braiding length of 140 and an error of $2.7\times 10^{-3}$ or $1.8\times 10^{-3}$ when the total charge of the five braiding anyons is $\mathbf{I}$ or $\tau$, respectively. The leakage error, defined by the largest transition amplitude from a computational to non-computational state, is 0 when the total charge of the braiding anyons is $\mathbf{I}$ and is $1.4\times 10^{-3}$ when the total charge is $\tau$.  We evaluate the approximation accuracy with the 2-norm of the difference between the target and approximated gate.

In general, the approximation accuracy is very close irrespective of the total charge of the braiding anyons, and we plot them together as a single data point in Fig. \ref{fig:2qubit_acc_comparison}. Importantly, the sequence lengths by the reinforcement learning algorithm and the brute force search both show a nearly linear scaling and are much smaller than the Solovay-Kitaev algorithm, indicating that similar to the single-qubit case in the main text, the reinforcement learning algorithm also has an edge in two-qubit quantum compiling scenarios.

\end{document}